\documentclass[12pt]{JHEP3}
\pdfoutput=1
\usepackage{graphicx}
\usepackage{amsmath}

\def\jpsi{{J/\psi}}

\def\sps{{\bigl.^1\hspace{-1mm}S^{[8]}_0}}

\def\pj{{\bigl.^3\hspace{-1mm}P^{[8]}_J}}
\def\p0{{\bigl.^3\hspace{-1mm}P^{[8]}_0}}

\def\mopa{{\langle\mathcal{O}^\jpsi(\bigl.^1\hspace{-1mm}S_0^{[8]})\rangle}}

\def\mopc{{\langle\mathcal{O}^\jpsi(\bigl.^3\hspace{-1mm}P_0^{[8]})\rangle}}

\def\to{\rightarrow}

\def\bqa{\begin{eqnarray}}
\def\eqa{\end{eqnarray}}
\def\bc{\begin{center}}
\def\bc{\end{center}}

\def\be{\begin{equation}}
\def\ee{\end{equation}}
\def\bea{\begin{eqnarray}}
\def\eea{\end{eqnarray}}

\normalbaselines

\title{Initial state radiation effects in inclusive $\jpsi$ production at B factories}

\author{Hua-Sheng Shao\\
Department of Physics and State Key Laboratory of Nuclear
Physics and Technology, Peking University,
Beijing 100871, China\\
E-mail: \email{erdissshaw@gmail.com}}

\abstract{Based on Monte Carlo techniques, we analyze the initial state radiation (ISR) effects in prompt $\jpsi$ inclusive production at B factories. ISR enhances cross section $\sigma(e^-e^+\to\jpsi+gg+X)$ by about $15-25\%$, which is almost the same size as the QCD and relativistic correction. Moreover, ISR slightly changes $\sigma(e^-e^+\to\jpsi+c\bar{c}+X)$. The $\jpsi$ momentum spectrum in $e^-e^+\to\jpsi+gg+X$ and in $e^-e^+\to\jpsi+c\bar{c}+X$ is softer after the photon showering from the initial $e^{\pm}$ beam radiation. After combining the QCD,relativistic, and ISR corrections,a more precise theoretical result is obtained. The new result provides a more stringent constraint of the color-octet contribution to $\sigma(e^-e^+\to\jpsi+X_{\rm{non-}c\bar{c}})$.}


\begin{document}

\section{Introduction}
A suitable method for studying the interplay between perturbative QCD and non-perturbative QCD is to investigate the quarkonium production at various colliders. The prompt $\jpsi$ production in hadronic collisions indicates that  color-octet mechanism plays a crucial role in reducing the discrepancies between theory and experiments~\cite{Butenschoen:2010rq,Butenschoen:2011yh,Ma:2010yw,Ma:2010jj,Chao:2012iv,
Butenschoen:2012px,Gong:2012ug}. Although color-singlet contribution may receive large QCD corrections in several examples~\cite{Artoisenet:2008fc,Lansberg:2006dh,Lansberg:2008gk,Lansberg:2013qka,Lansberg:2011hi,Lansberg:2009db}, difficulty in explaining the simultaneous yields and polarization of prompt  $\jpsi$ in hadronic collisions is still encountered.A significant color-octet component is important to explain the high-transverse momentum hadronic data in the current framework. By contrast, the color-singlet one in a relatively small-scale physics regime is sufficient to explain the experimental data at hadron colliders~\cite{Maltoni:2006yp,Brodsky:2009cf} and B factories~\cite{Zhang:2006ay,Gong:2009ng,Ma:2008gq,Gong:2009kp,He:2009uf,Jia:2009np,Zhang:2009ym}. Hence, the phyiscs of these two scales should be studied further. 

In this research, we consider the prompt $\jpsi$ inclusive production in $e^-e^+$ collisions at center-of-mass energy $\sqrt{s}=10.6~\rm{GeV}$. 
Over a decade ago, \textbf{BABAR} collaboration and \textbf{BELLE} collaboration reported the cross section $\sigma(e^-e^+\to \jpsi+X)$ is $2.52\pm0.21\pm0.21 ~\rm{pb}$~\cite{Aubert:2001pd} and  $1.47\pm0.10\pm0.13~\rm{pb}$~\cite{Abe:2001za} respectively. By contrast,the leading order (LO) color-singlet theoretical prediction~\cite{Yuan:1996ep,Cho:1996cg,Baek:1998yf,Schuler:1998az,Kiselev:1994pu,Liu:2003zr} that mainly includes the contributions from the processes $e^-e^+\to \jpsi+c\bar{c}+X$ and $e^-e^+\to \jpsi+gg+X$, was at least $3-5$ times lower than the obtained measurements. This finding suggest a substantial color-octet contribution that can be generated at lower $\alpha_s$ power via $e^-e^+\to \jpsi(\pj,\sps)+g$. Moreover, the \textbf{BELLE} collaboration measured the associated production cross section $\sigma(e^-e^+\to\jpsi+c\bar{c}+X)=0.87^{+0.21}_{-0.19}\pm0.17~\rm{pb}$ and the ratio
 $R_{c\bar{c}}=\frac{\sigma(e^-e^+\to\jpsi+c\bar{c}+X)}{\sigma(e^-e^+\to\jpsi+X)}=0.59^{+0.15}_{-0.13}\pm0.12$~\cite{Abe:2002rb}. The cross section for $\jpsi+c\bar{c}+X$ is at least a factor of five and higher than LO color-singlet~\cite{Cho:1996cg,Baek:1998yf,Schuler:1998az,Kiselev:1994pu,Liu:2003zr,Liu:2003jj} and color-octet~\cite{Liu:2003jj} theoretical estimations. Several theoretical improvements are formulated subsequently to decrease the large discrepancies~\cite{Fleming:2003gt,Lin:2004eu,Leibovich:2007vr}. Specifically, next-to-leading order (NLO) QCD correction is used to obtain a large enhancement~\cite{Zhang:2006ay} of the cross section for $\jpsi+c\bar{c}+X$, which has been confirmed by other authors~\cite{Gong:2009ng}. The result is comparable with the experimental data in some specific parameter choices. Moreover, the QCD~\cite{Ma:2008gq,Gong:2009kp} and relativistic correction~\cite{He:2009uf,Jia:2009np} enhance the color-singlet cross section $\sigma(e^-e^+\to\jpsi+gg+X)$ by about $20-30\%$ respectively. The QCD correction~\cite{Ma:2008gq,Gong:2009kp} to $e^-e^+\to\jpsi+gg+X$ significantly changes the  $\jpsi$ momentum distribution,particularly near the kinematic endpoint.This changes is important because the color-octet part~\cite{Braaten:1995ez} enhances the cross section at the endpoint prior to resummation~\cite{Fleming:2003gt}\footnote{Resummation is heavily based on a phenomenological shape function~\cite{Fleming:2003gt}.}. Thus, the $\jpsi$ momentum spectrum might provide a constraint to color-octet matrix elements. A comparison of the theoretical result with the current \textbf{BELLE} measurements~\cite{Pakhlov:2009nj} was performed in ref.\cite{Zhang:2009ym}. A strong constraint on the non-perturbative color-octet matrix elements was extracted as~\cite{Zhang:2009ym}
\bqa
\mopa+4.0\frac{\mopc}{m_c^2}<(2.0\pm0.6)\times 10^{-2}~\rm{GeV}^3,
\eqa
which apparently contradicts the high $p_T$ hadronic results in non-relativistic QCD (NRQCD) theory~\cite{Bodwin:1994jh} at NLO level~\cite{Ma:2010yw,Ma:2010jj,Chao:2012iv,Gong:2012ug,Bodwin:2014gia,Faccioli:2014cqa}\footnote{In refs.\cite{Butenschoen:2010rq,Butenschoen:2011yh,Butenschoen:2012px}, the authors obtained a color-octet matrix elements set satisfying this constraint by using the world yields data. However, in their fit, they included many small transverse momentum data. Hence, they predicted a completely wrong $\jpsi$ polarization at Tevatron and LHC~\cite{Butenschoen:2012px}.}. 

\begin{center}
\begin{figure}
\hspace{0cm}\includegraphics[width=0.45\textwidth]{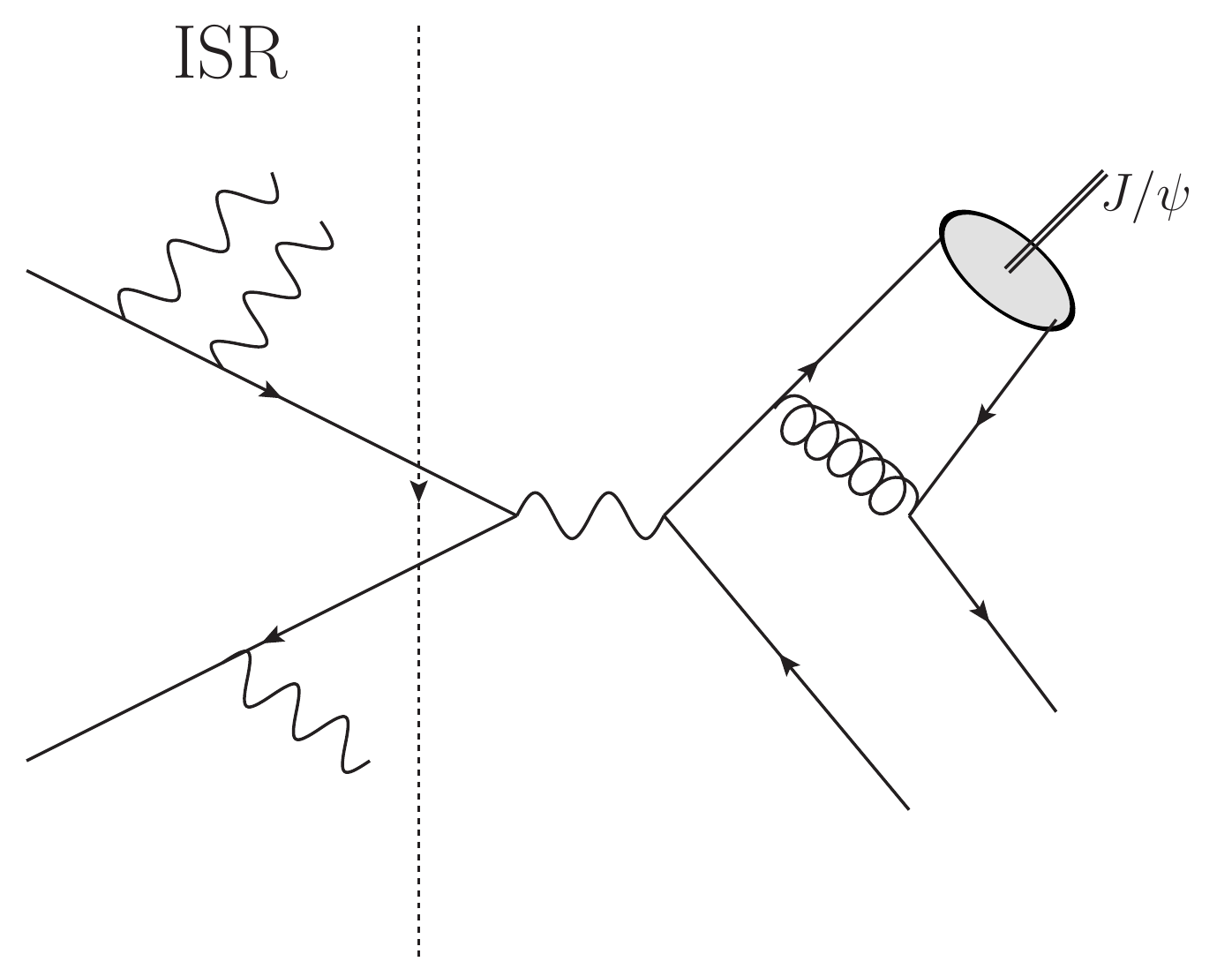}
\hspace{0cm}\includegraphics[width=0.45\textwidth]{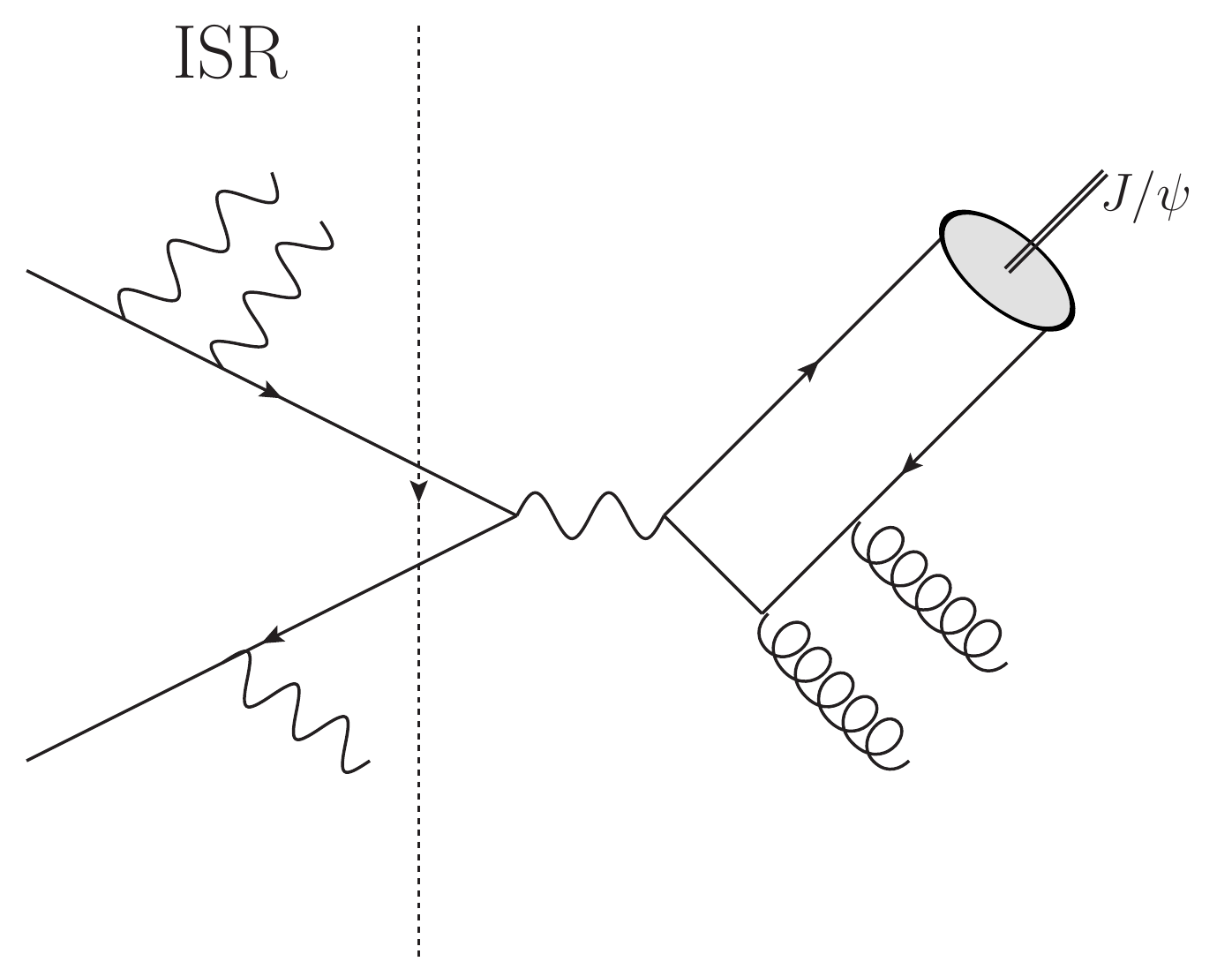}
\caption{\label{fig:FeynDiag}Two representative Feynman diagrams with ISR.}
\end{figure}
\end{center}

\begin{center}
\begin{figure}
\hspace{0cm}\includegraphics[width=0.9\textwidth]{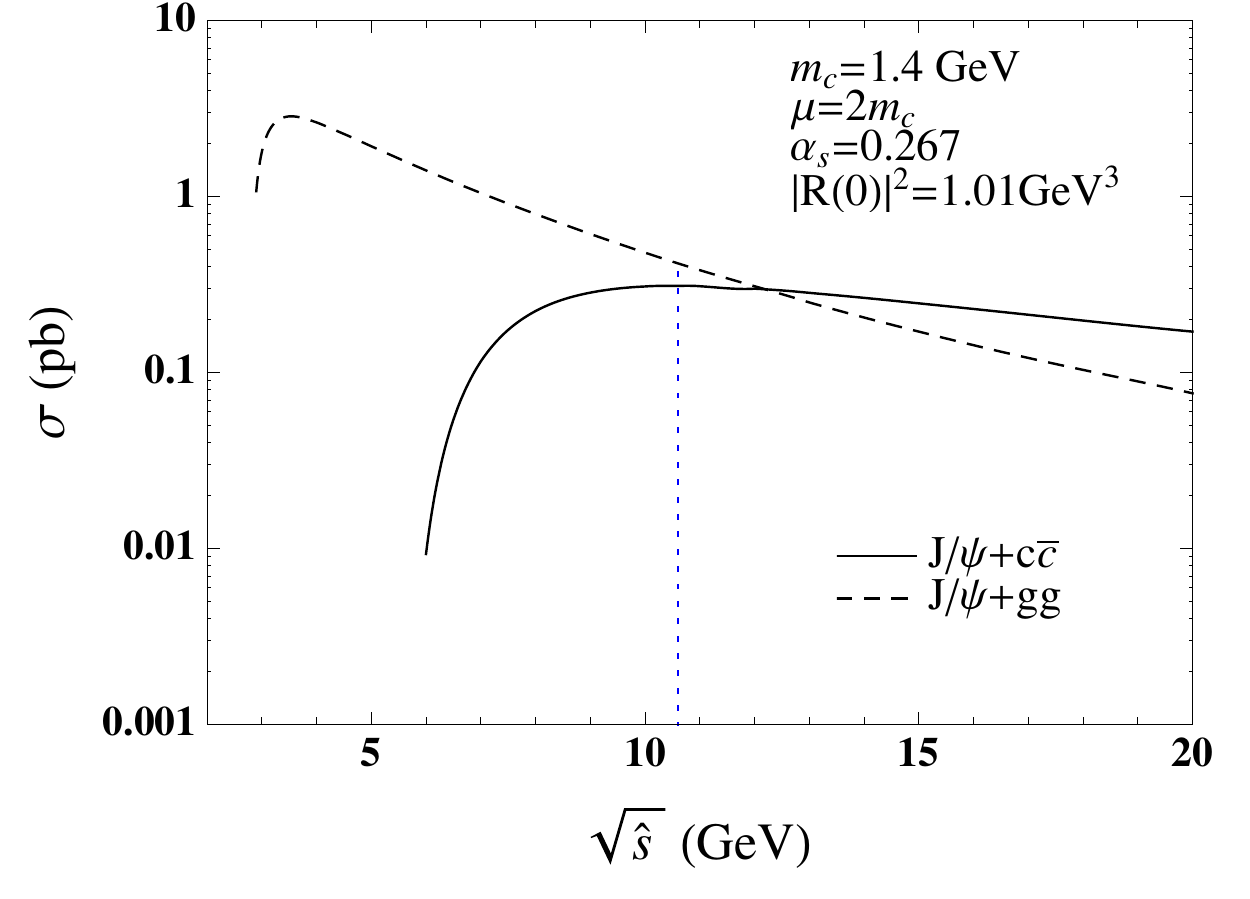}
\caption{\label{fig:dsigmads}Cross sections  as functions of center-of-mass energy $\sqrt{\hat{s}}$ in hard reaction.}
\end{figure}
\end{center}

In this article, we examine the initial state radiation (ISR) effect to prompt $\jpsi$ inclusive production at B factories. ISR is an extremely important feature that should be considered in investigating the physics in electron-positron collisions. After including the ISR effect in the inclusive $\jpsi$ production at B factories, we obtain a more precise theoretical result. In general, the detailed studies of photon-radiative corrections from the initial $e^{\pm}$ beams require Monte Carlo (MC) generators~\cite{Bonvicini:1988vv,Jadach:1988gb,Jadach:1991by,Dahmen:1990cd,Caffo:1991cg,
Caffo:1993hc,Fujimoto:1993qh,Fujimoto:1993ge,Munehisa:1995si}. We interface a general-purpose matrix element and events generator \textbf{HELAC-Onia}~\cite{Shao:2012iz,Kanaki:2000ms,Kanaki:2000ey,Papadopoulos:2006mh,Cafarella:2007pc} to the general photon shower program \textbf{QEDPS}~\cite{Fujimoto:1993qh,Fujimoto:1993ge,Munehisa:1995si} to include the ISR in various $e^-e^+$ annihilation processes. In the two processes $e^-e^+\to \jpsi+c\bar{c}+X$ and $e^-e^+\to \jpsi+gg+X$, two representative Feynman diagrams with ISR are shown in figure~\ref{fig:FeynDiag}. After the photon shower, the annihilating $e^-$ and $e^+$ do not produce head-on collisions because they might deviate from the beam axis through the radiation. The center-of-mass energy after showering $\sqrt{\hat{s}}$ is lower than $\sqrt{s}=10.6~\rm{GeV}$. As shown in figure~\ref{fig:dsigmads}, the LO cross section for $e^-e^+\to \jpsi+gg+X$ increases as $\sqrt{\hat{s}}$ decreases near $10.6~\rm{GeV}$, whereas that for $e^-e^+\to \jpsi+c\bar{c}+X$ changes minimally as $\sqrt{\hat{s}}$ decreases near $10.6~\rm{GeV}$. Therefore, the ISR effect should be significant in $\jpsi+gg+X$ , whereas it might not be that important in $\jpsi+c\bar{c}+X$.Considering that ISR and $\jpsi$ production are factorizable, ISR should not change the QCD correction and relativistic correction unless the K factors of these two corrections considerably vary with $\sqrt{\hat{s}}$ near $10.6~\rm{GeV}$. We will study the ISR effect in detail by including the QCD correction and relativistic correction.

The article is organized as follows: we will study the ISR effects in $e^-e^+\to \jpsi+c\bar{c}+X$ in section \ref{sec:2} and in $e^-e^+\to \jpsi+gg+X$ in section \ref{sec:3}. We then formulate our conclusion in section \ref{sec:4}.

\section{ISR in $\jpsi c\bar{c}+X$\label{sec:2}}
The inclusive double charm production at B factories is one of the most interesting processes for probing heavy quarkonium physics. The most precise measurement of its cross section performed by \textbf{BELLE} collaboration is~\cite{Pakhlov:2009nj}
\bqa
\sigma_{\rm{prompt}}(e^-e^+\to \jpsi+c\bar{c}+X)=0.74\pm0.08^{+0.09}_{-0.08}~\rm{pb}.
\eqa
The color-singlet cross section with NLO QCD correction is $0.33(0.47)~\rm{pb}$ when $m_c=1.5(1.4)~\rm{GeV}$ and $\mu=2m_c$,$|R(0)|^2=1.01~\rm{GeV}^3$. The QED and double photon contributions enhance the cross section by $8+29~\rm{fb}$~\cite{Zhang:2009ym}. The feed-down contribution from $\psi(2S)$ increases the cross section by a factor of $1.355$, whereas that from $\chi_{cJ}$ is $21~\rm{fb}$~\cite{Zhang:2009ym,Liu:2003jj}. The small color-octet contribution is $11~\rm{fb}$~\cite{Liu:2003jj}. After these combinations are completed, the prompt cross section becomes $0.51(0.71)~\rm{pb}$~\cite{Zhang:2009ym}. The relativistic correction was performed in ref.\cite{He:2007te}, but its effect was negligible.

\begin{center}
\begin{figure}
\hspace{0cm}\includegraphics[width=0.45\textwidth]{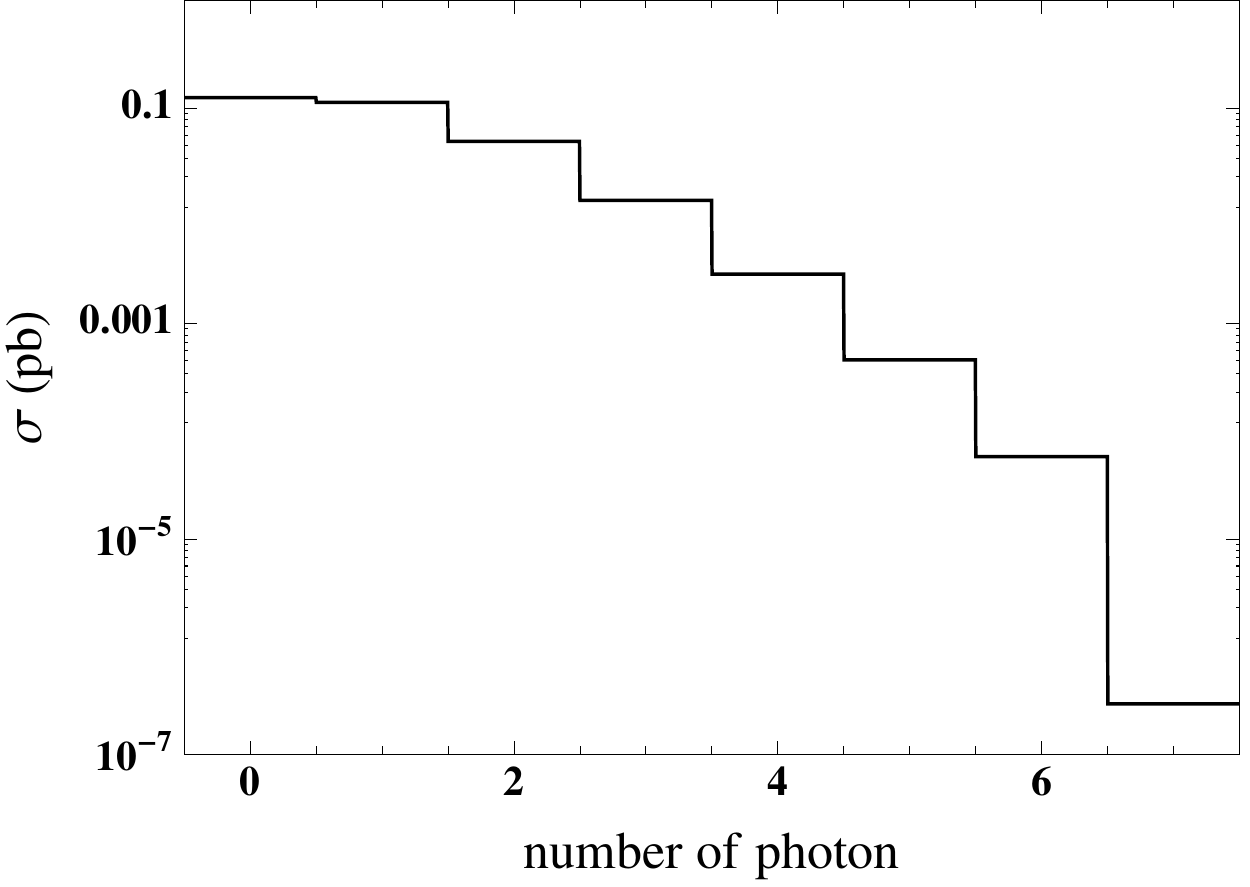}
\hspace{0cm}\includegraphics[width=0.45\textwidth]{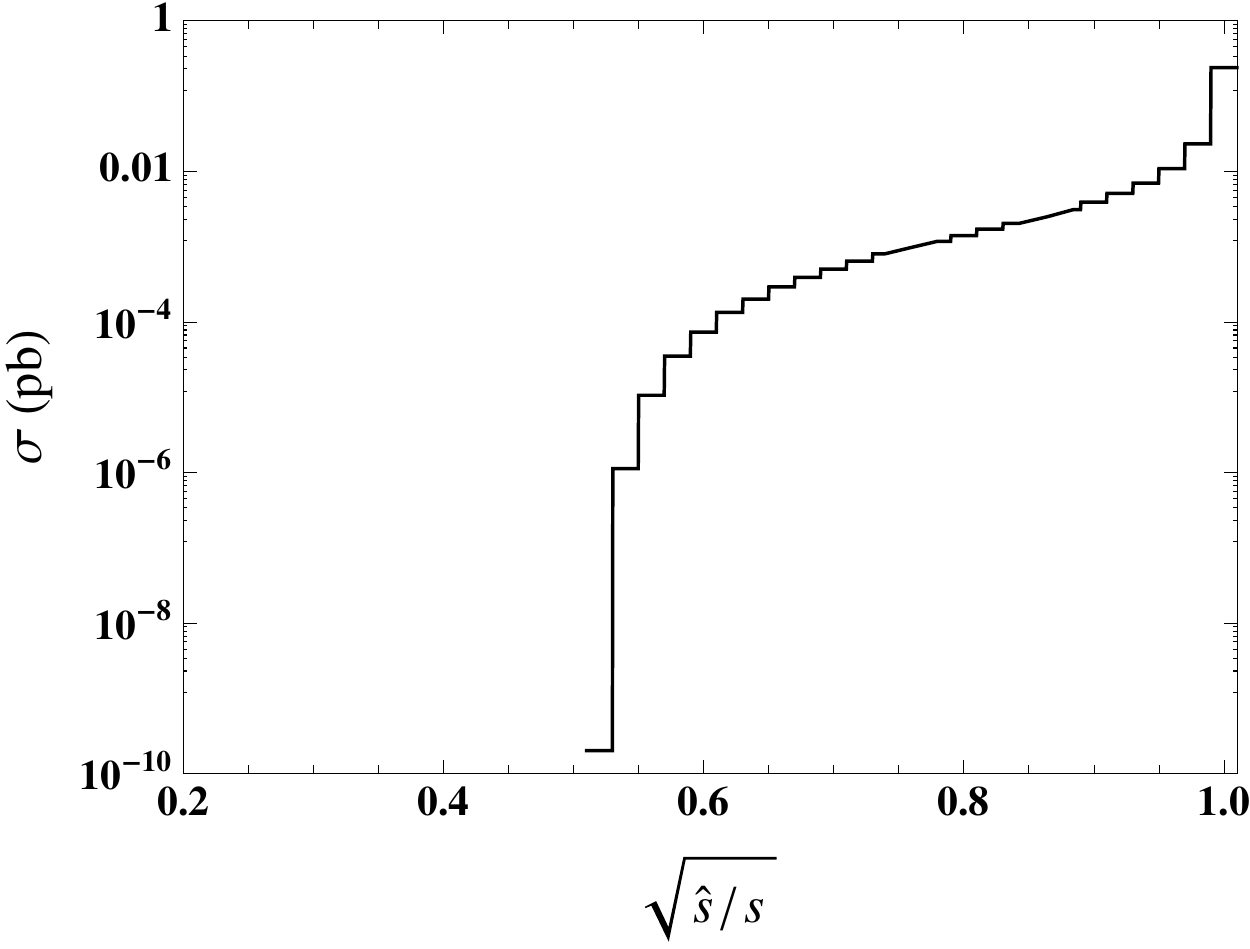}
\caption{\label{fig:NAratioscc}Cross section as functions of number of photon and $\sqrt{\hat{s}}/\sqrt{s}$ in $e^-e^+\to \jpsi+c\bar{c}+X$.}
\end{figure}
\end{center}

However, the physical cross section should always include ISR at $e^-e^+$ annihilation. Figure~\ref{fig:NAratioscc} shows the cross sections with a number of radiated photons for $e^-e^+\to \jpsi+c\bar{c}+X$. A substantial probability exists to radiate at least a photon. The average number of ISR photon in each event is about $0.88$, which can be presumed because besides an $\alpha$ suppression for radiating a photon, there is also an extra $\log(m_e/\sqrt{s})$ enhancement.  To determine the changes of the center-of-mass energy $\sqrt{\hat{s}}$ after showering, we also plot the $\sqrt{\hat{s}}/\sqrt{s}$ distribution in figure~\ref{fig:NAratioscc}. The average value for $\sqrt{\hat{s}}/\sqrt{s}$ is about $0.98$. The value is quite close to $1$, which indicates that ISR correction is small.

\begin{center}
\begin{figure}
\hspace{0cm}\includegraphics[width=0.45\textwidth]{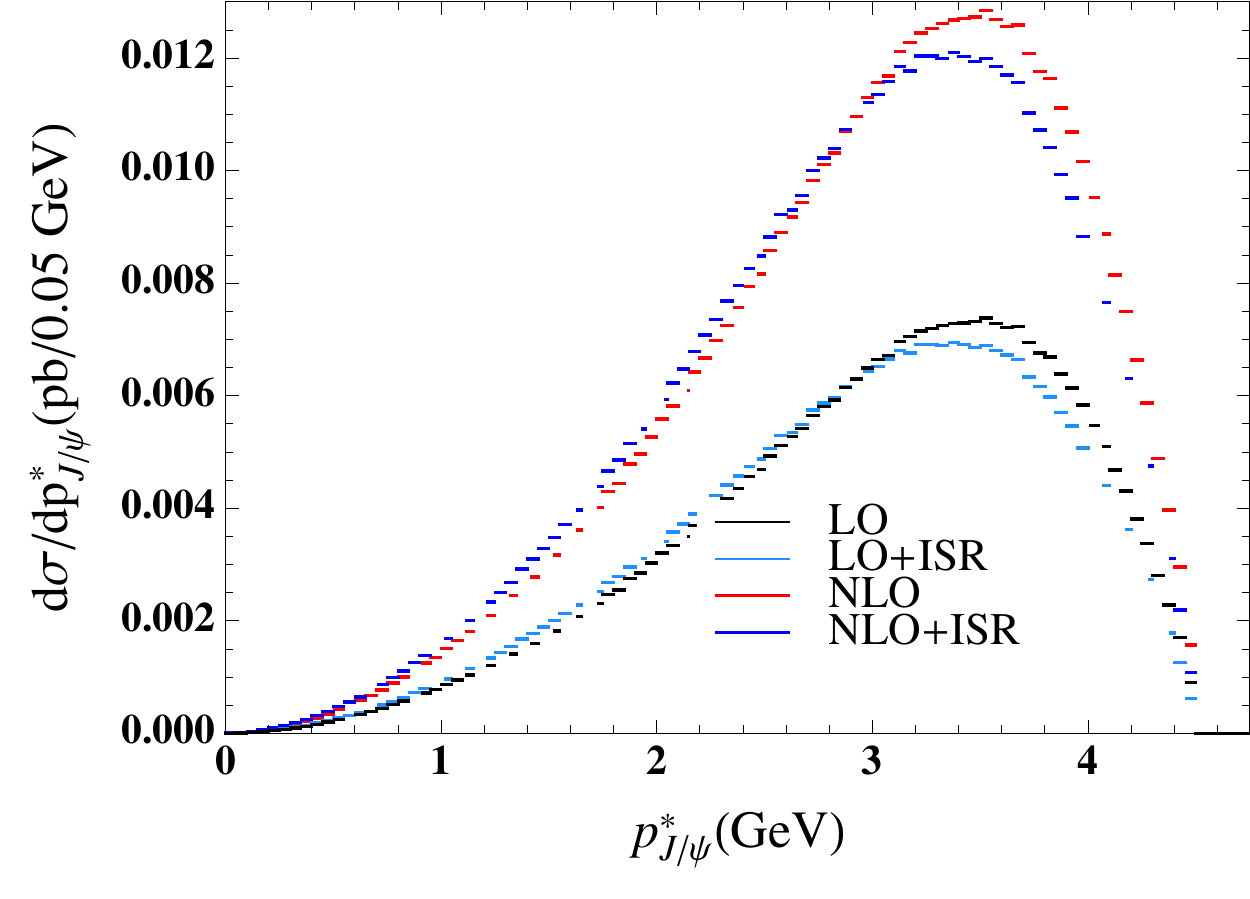}
\hspace{0cm}\includegraphics[width=0.45\textwidth]{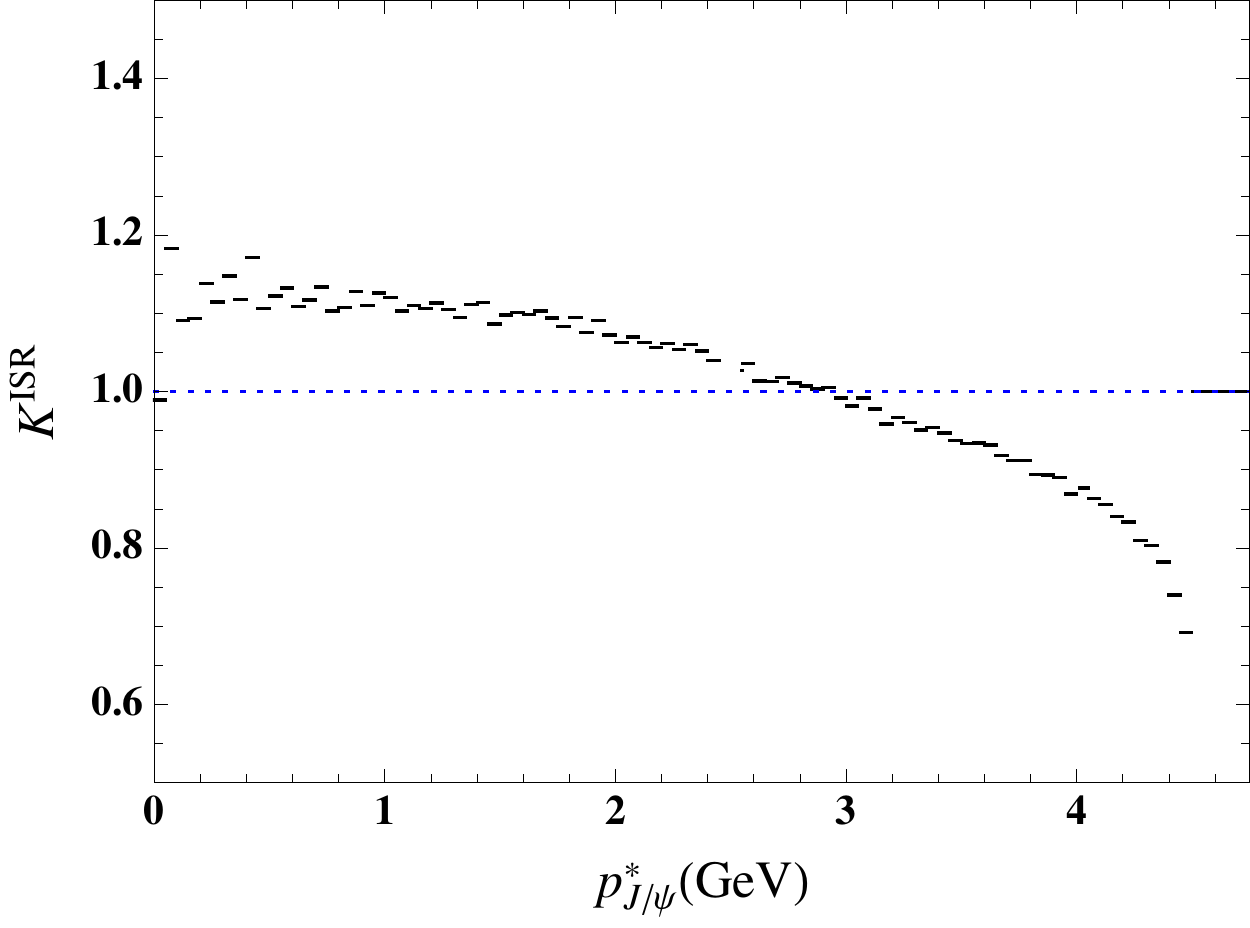}
\caption{\label{fig:dsigmadpcc}Cross sections and $K^{\rm{ISR}}=\sigma^{\rm{LO+ISR}}/\sigma^{\rm{LO}}$ as functions of the $\jpsi$ momentum $p^*_{\jpsi}$ in the  rest frame of initial $e^-e^+$.We take the parameter set as $m_c=1.4~\rm{GeV},\mu=2m_c$  in $e^-e^+\to \jpsi+c\bar{c}+X$.}
\end{figure}
\end{center}
\begin{center}
\begin{figure}
\hspace{0cm}\includegraphics[width=0.45\textwidth]{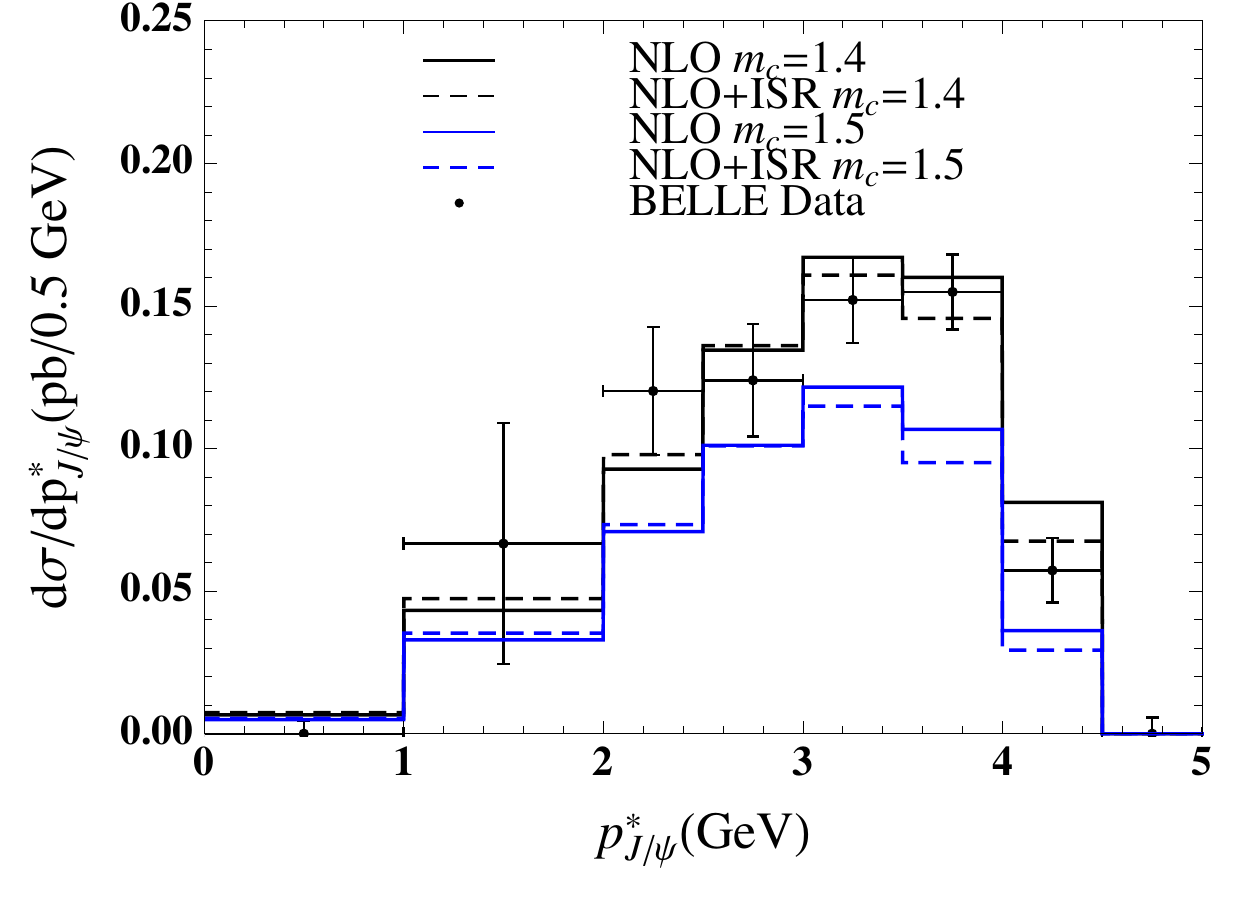}
\hspace{0cm}\includegraphics[width=0.45\textwidth]{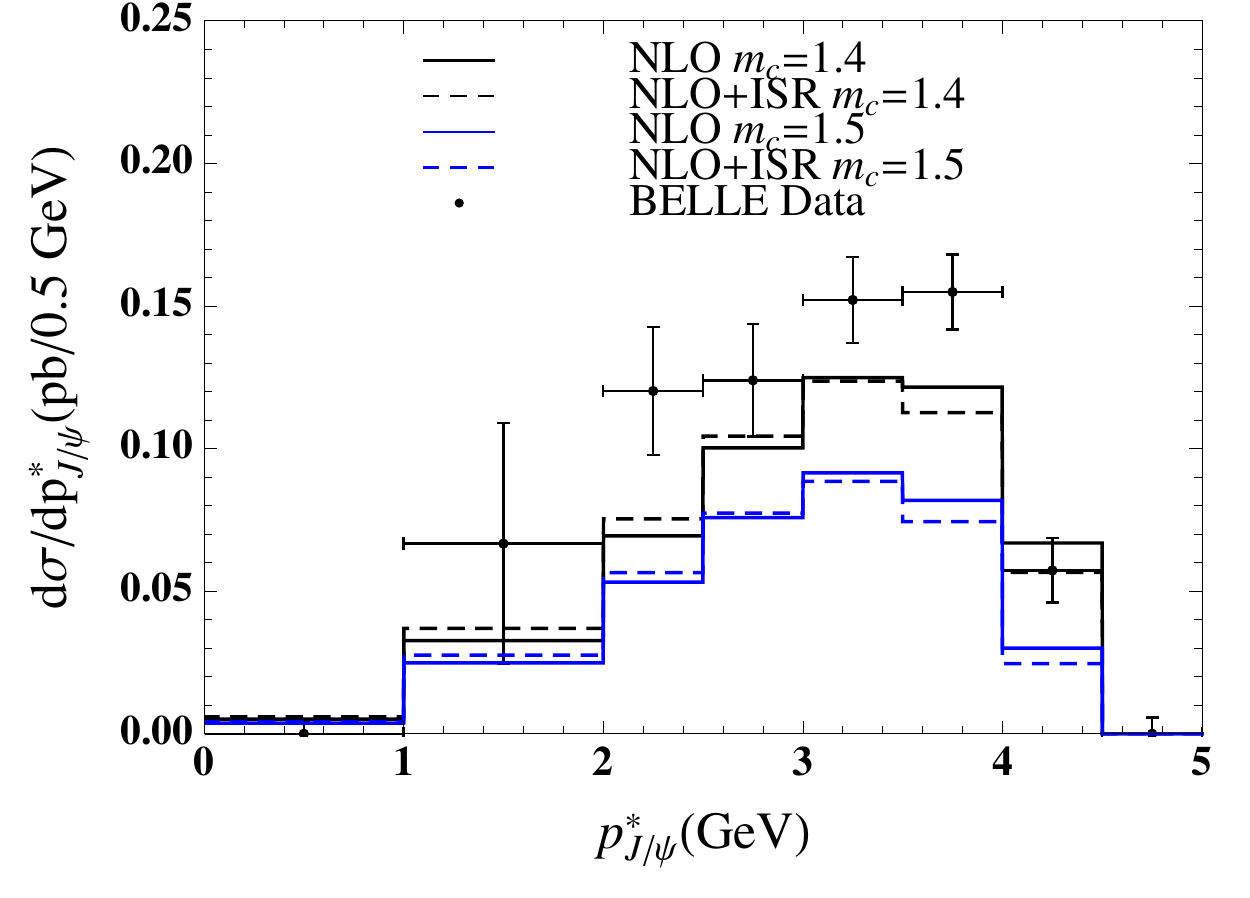}
\caption{\label{fig:dsigmadpBELLEcc}Comparisons of the theoretical predictions and \textbf{BELLE} measurement~\cite{Pakhlov:2009nj} with $\mu=2m_c$ (left-panel) and $\mu=\sqrt{\hat{s}}/2$ (right-panel) respectively in $e^-e^+\to \jpsi+c\bar{c}+X$. We have multiplied a factor 1.355 to account for the feed-down contribution from $\psi(2S)\to \jpsi+X$.}
\end{figure}
\end{center}

Figure~\ref{fig:dsigmadpcc} shows the ISR effect in the $\jpsi$ momentum spectrum. We obtain the curves for $\rm{NLO}$ and $\rm{NLO+ISR}$ by normalizing the corresponding $\rm{LO}$ and $\rm{LO+ISR}$\footnote{$\rm{LO+ISR}$ refers to LO with ISR effects. LO result already includes the QCD and QED diagrams.} results by a NLO K factor from ref.\cite{Zhang:2009ym}, because the K factor changes slightly in the $p_{\rm{\jpsi}}^{*}$\footnote{$p_{\rm{\jpsi}}^{*}$ refers to the momentum of $\jpsi$ in the rest frame of the initial $e^-e^+$ before showering.} spectrum and with $\sqrt{\hat{s}}$~\cite{Gong:2009ng}. We take $m_c=1.4~\rm{GeV},\mu=2m_c,|R(0)|^2=1.01~\rm{GeV}^3$ here. As expected, ISR minimally corrects the momentum spectrum. We also present $K^{\rm{ISR}}=\sigma^{\rm{LO+ISR}}/\sigma^{\rm{LO}}$ as a function of $p_{\rm{\jpsi}}^{*}$ in the right panel of figure~\ref{fig:dsigmadpcc} for further clarification. ISR causes the $\jpsi$ momentum spectrum to become slightly softer. We also compare the \textbf{BELLE} measurement~\cite{Pakhlov:2009nj} with the theoretical prompt results in figure~\ref{fig:dsigmadpBELLEcc}. Considering the uncertainties in the input parameters, such as $m_c,\mu$, we compare the experimental with the theoretical results in different parameter sets. It is shown that $m_c=1.4~\rm{GeV},\mu=2m_c$ is the closest set to the \textbf{BELLE} data~\cite{Pakhlov:2009nj}, although large uncertainties remain in the experimental data. Finally, the prompt total cross sections are summarized in table~\ref{tab:sigmacc}. ISR decreases the cross section by an extremely minimal amount.

\begin{table}
\begin{center}
\begin{tabular}{{c}*{4}{c}}\hline\hline
parameter sets & LO (pb) & LO+ISR (pb) & NLO (pb) & NLO+ISR (pb)\\\hline $m_c=1.4~\rm{GeV},\mu=2m_c$ & $0.45$ & $0.44$ & $0.77$ & $0.75$ \\
$m_c=1.5~\rm{GeV},\mu=2m_c$ & $0.31$ & $0.30$ & $0.54$ & $0.53$ \\
$m_c=1.4~\rm{GeV},\mu=\sqrt{\hat{s}}/2$ & $0.31$ & $0.31$ & $0.59$ & $0.59$ \\
$m_c=1.5~\rm{GeV},\mu=\sqrt{\hat{s}}/2$ & $0.23$ & $0.22$ & $0.42$ & $0.42$
\\\hline\hline
\end{tabular}
\end{center}
\caption{\label{tab:sigmacc}Cross sections of $e^-e^+\to \jpsi+c\bar{c}+X$ in different parameter sets.}
\end{table}

\section{ISR in $\jpsi gg+X$\label{sec:3}}
In this section, we will study the process $e^-e^+\to \jpsi+gg+X$. \textbf{BELLE} collaboration has measured the cross section for $\jpsi+X_{\rm{non-}c\bar{c}}$ in ref.\cite{Pakhlov:2009nj} as
\bqa
\sigma_{\rm{prompt}}(e^+e^-\to \jpsi+X_{\rm{non-}c\bar{c}})=0.43\pm0.09\pm0.09~\rm{pb}.
\eqa
Theoretically, the NLO color-singlet cross section for the prompt $\jpsi+gg+X$ is $0.67(0.53)~\rm{pb}$ when $m_c=1.4~\rm{GeV},\mu=2.8(5.3)~\rm{GeV},|R(0)|^2=1.01~\rm{GeV}^3$~\cite{Ma:2008gq,Gong:2009kp}, which enhances the LO cross section by about $20-30\%$. Subsequently, the authors of ref.\cite{He:2009uf} found that the relativistic correction also contributes a factor of $20-30\%$ to the color-singlet $\sigma(e^+e^-\to \jpsi+gg+X)$, which is comparable to that of the QCD correction. The relativistic correction was also confirmed by other authors~\cite{Jia:2009np}. Based on their calculations, the color-singlet result is already saturating the experimental data upon the assumption that the whole partonic cross section contributes to $\jpsi+X_{\rm{non-}c\bar{c}}$ final states.

\begin{center}
\begin{figure}
\hspace{0cm}\includegraphics[width=0.45\textwidth]{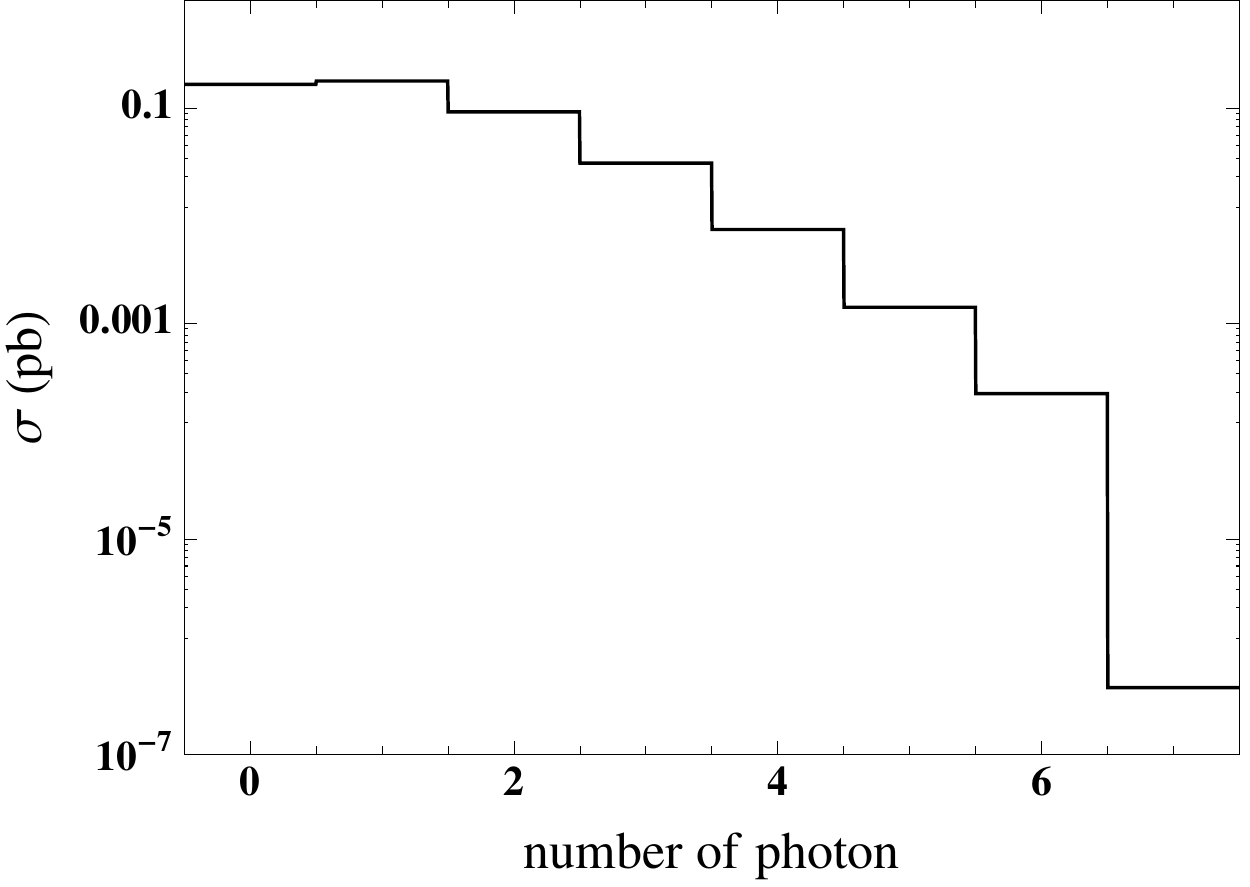}
\hspace{0cm}\includegraphics[width=0.45\textwidth]{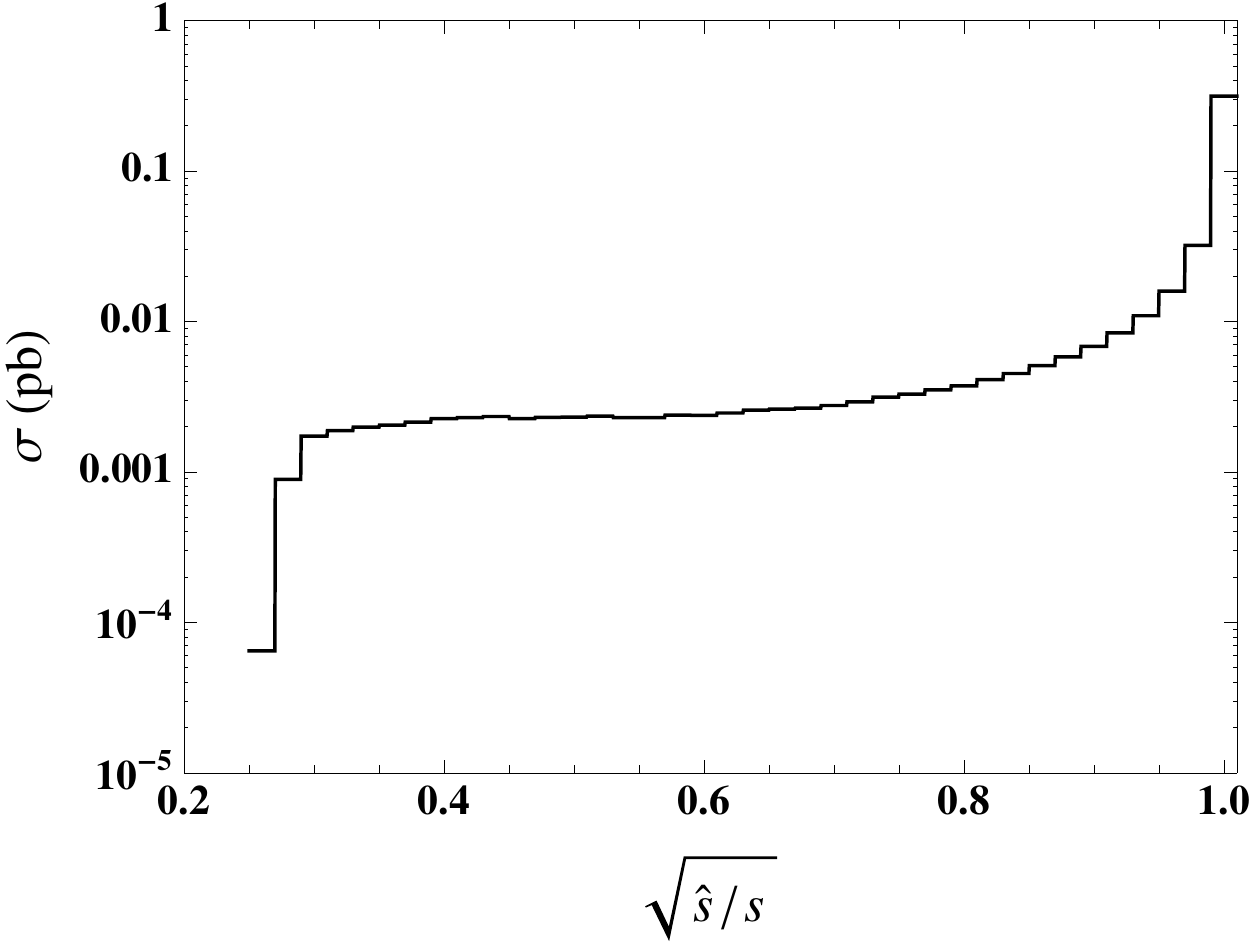}
\caption{\label{fig:NAratios}Cross section as functions of number of photon and $\sqrt{\hat{s}}/\sqrt{s}$ in $e^-e^+\to \jpsi+gg+X$.}
\end{figure}
\end{center}

As discussed in the first section, ISR should significantly change the cross section. We plot the number of ISR photon distribution and $\sqrt{\hat{s}}/\sqrt{s}$ in figure~\ref{fig:NAratios}. Compared with $\jpsi+c\bar{c}+X$, a slightly higher probability to radiate a photon is observed in $\jpsi+gg+X$. The reason mainly relies on the cross section $\sigma(e^-e^+\to\jpsi+gg+X)$ that increases as $\sqrt{\hat{s}}$ decreases. The average number of photon per event improves to $1.04$. Moreover, the average $\sqrt{\hat{s}}/\sqrt{s}$ is $0.93$. Therefore, the ISR effects are more important in $e^-e^+\to \jpsi+gg+X$.

\begin{center}
\begin{figure}
\hspace{0cm}\includegraphics[width=0.45\textwidth]{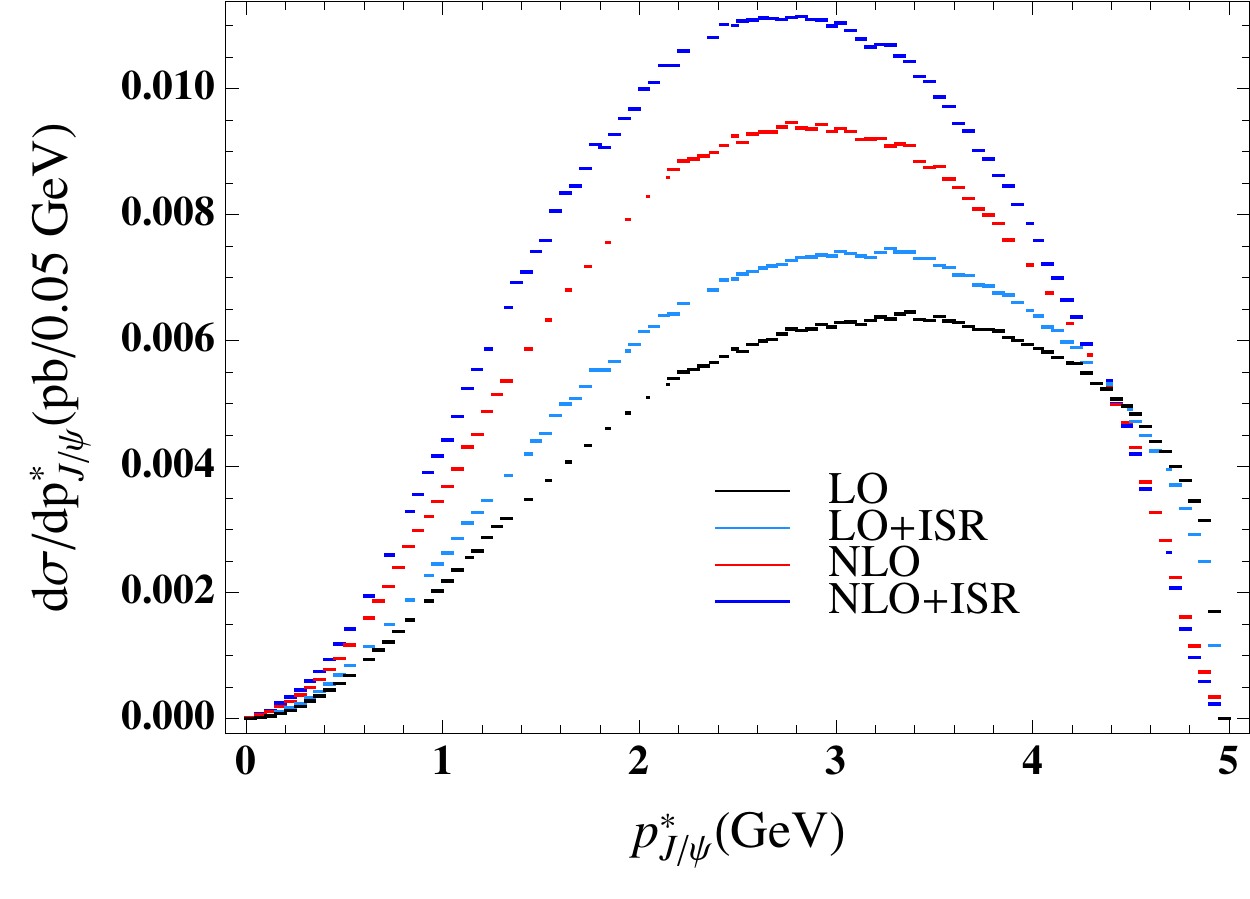}
\hspace{0cm}\includegraphics[width=0.45\textwidth]{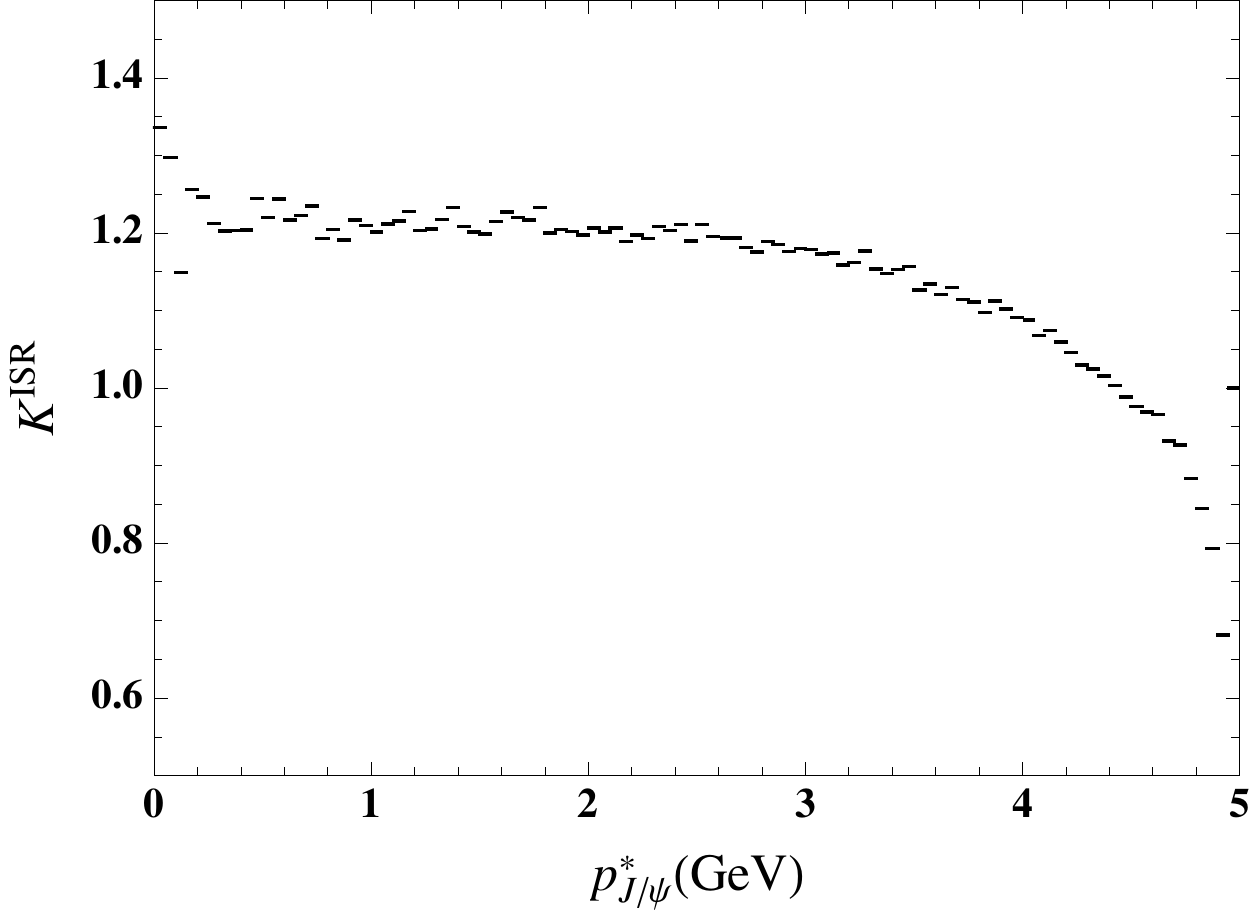}
\caption{\label{fig:dsigmadp}Cross sections and $K^{\rm{ISR}}=\sigma^{\rm{LO+ISR}}/\sigma^{\rm{LO}}$ as functions of the $\jpsi$ momentum $p^*_{\jpsi}$ in the rest frame of initial $e^-e^+$.We take the parameter set as $m_c=1.4~\rm{GeV},\mu=2m_c$  in $e^-e^+\to \jpsi+gg+X$.}
\end{figure}
\end{center}
\begin{center}
\begin{figure}
\hspace{0cm}\includegraphics[width=0.45\textwidth]{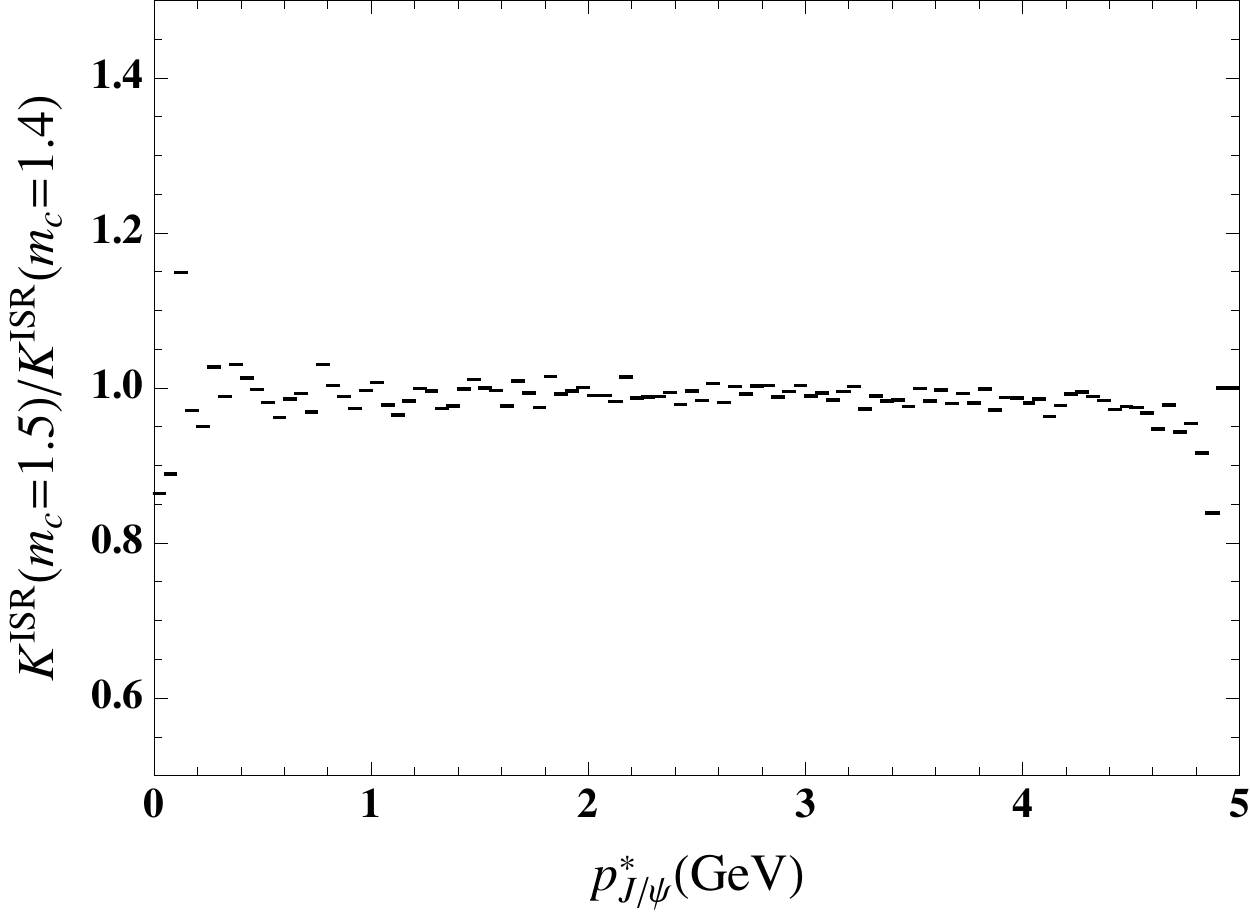}
\hspace{0cm}\includegraphics[width=0.45\textwidth]{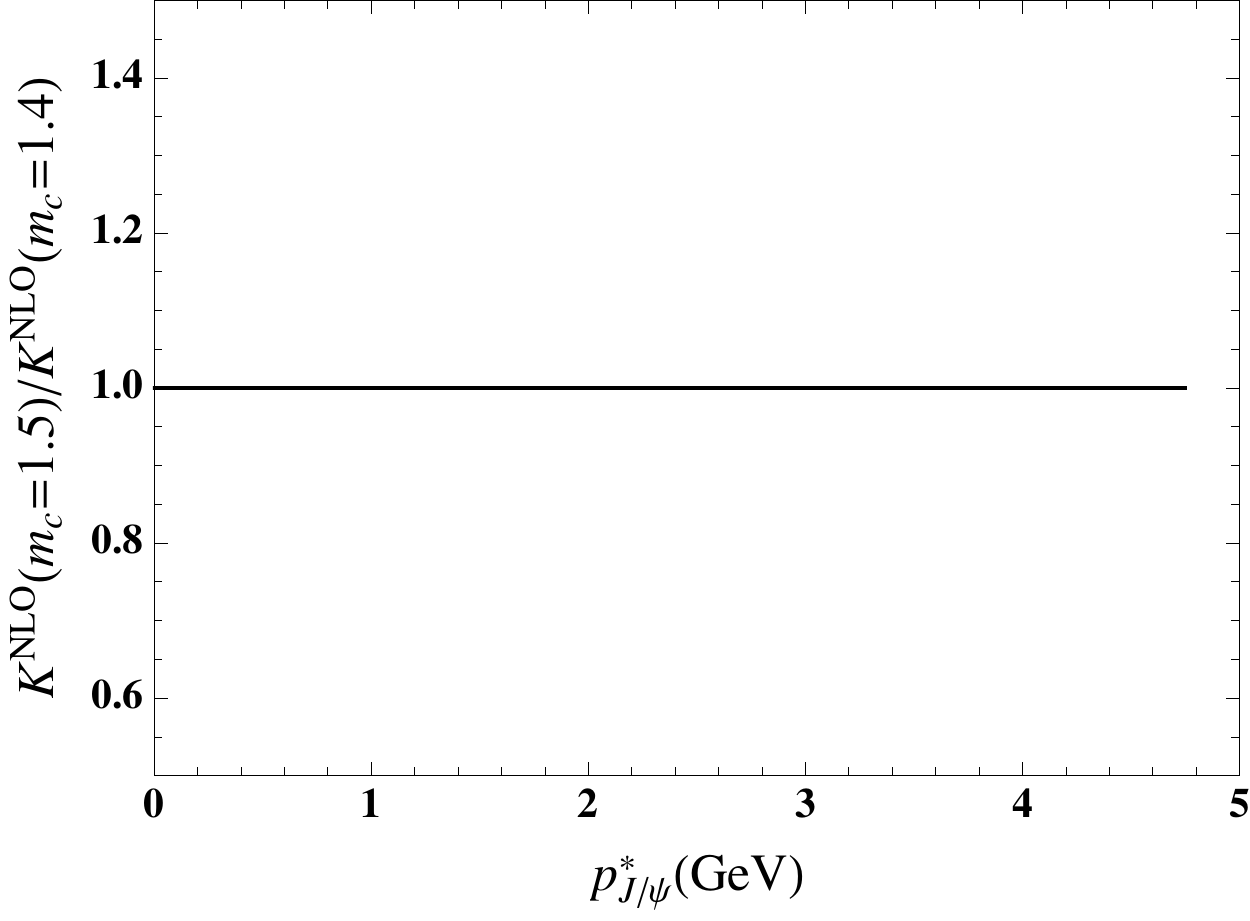}
\caption{\label{fig:KISR1514}$K^{\rm{ISR}}=\sigma^{\rm{LO+ISR}}/\sigma^{\rm{LO}}$ (left) and $K^{\rm{NLO}}=\sigma^{\rm{NLO}}/\sigma^{\rm{LO}}$ (right) as functions of $\jpsi$ momentum $p^*_{\jpsi}$ with $m_c=1.5~\rm{GeV}$ and $m_c=1.4~\rm{GeV}$ and $\mu=2m_c$  in $e^-e^+\to \jpsi+gg+X$ than in $e^-e^+\to\jpsi+c\bar{c}+X$.}
\end{figure}
\end{center}

Next, we intend to include the QCD correction,relativistic correction and ISR in the color-singlet results. By contrast to the case in $\jpsi+c\bar{c}$, the QCD correction to $\jpsi+gg$ results in $\jpsi$ softer momentum spectrum than the LO one~\cite{Ma:2008gq}. At the endpoint, the LO result suffers from the large logarithms $\log(1-E_{\jpsi}/E^{\rm{max}}_{\jpsi})$ because of kinematic reasons. The LO spectrum significantly changes at the endpoint with the resummation of such logarithms, whereas resummation has limited effects on the NLO spectrum~\cite{Ma:2008gq}\footnote{Resummation also has a very minimal effect on the total cross section~\cite{Ma:2008gq}.}. Moreover, the relativistic correction should not change the LO spectrum but enhance it by a simple K factor. We use the formula
\bqa
\frac{d\sigma^{\rm{NLO}}}{dp_{\jpsi}^*}&=&(K^{\rm{NLO}(\alpha_s)}+K^{\rm{NLO}(v^2)}-1)\frac{d\sigma^{\rm{LO}}}{dp_{\jpsi}^*}\nonumber\\
K^{\rm{NLO}(\alpha_s)}&=&\frac{d\sigma^{\rm{NLO}(\alpha_s)}}{dp_{\jpsi}^*}/\frac{d\sigma^{\rm{LO}}}{dp_{\jpsi}^*},\nonumber\\
K^{\rm{NLO}(v^2)}&=&\frac{d\sigma^{\rm{NLO}(v^2)}}{dp_{\jpsi}^*}/\frac{d\sigma^{\rm{LO}}}{dp_{\jpsi}^*},
\eqa
to obtain the fixed-order result by considering the QCD and relativistic corrections. A similar formula can be applied to the result with ISR
\bqa
\frac{d\sigma^{\rm{NLO+ISR}}}{dp_{\jpsi}^*}&=&(K^{\rm{NLO}(\alpha_s)}+K^{\rm{NLO}(v^2)}-1)\frac{d\sigma^{\rm{LO+ISR}}}{dp_{\jpsi}^*}.
\eqa
This treatment is justified because the ISR and QCD/relativistic correction can be factorized. Moreover, the K factor of the QCD/relativistic correction slightly changes with $\sqrt{\hat{s}}$~\cite{Gong:2009kp,He:2009uf}. The result is shown in figure~\ref{fig:dsigmadp}. ISR results in softer $\jpsi$ momentum spectrum, which is apparent from the $K^{\rm{ISR}}=\sigma^{\rm{LO+ISR}}/\sigma^{\rm{LO}}$ shown in the right panel of figure~\ref{fig:dsigmadp}. Another interesting aspect that we want to determine is the sensitivity of the K factors to $m_c$ values. We establish two plots in figure~\ref{fig:KISR1514}. As shown in the figure, the values of $K^{\rm{ISR}}=\sigma^{\rm{LO+ISR}}/\sigma^{\rm{LO}}$ and $K^{\rm{NLO}}=K^{\rm{NLO}(\alpha_s)}+K^{\rm{NLO}(v^2)}-1$ are insensitive to $m_c$.

\begin{center}
\begin{figure}
\hspace{0cm}\includegraphics[width=0.45\textwidth]{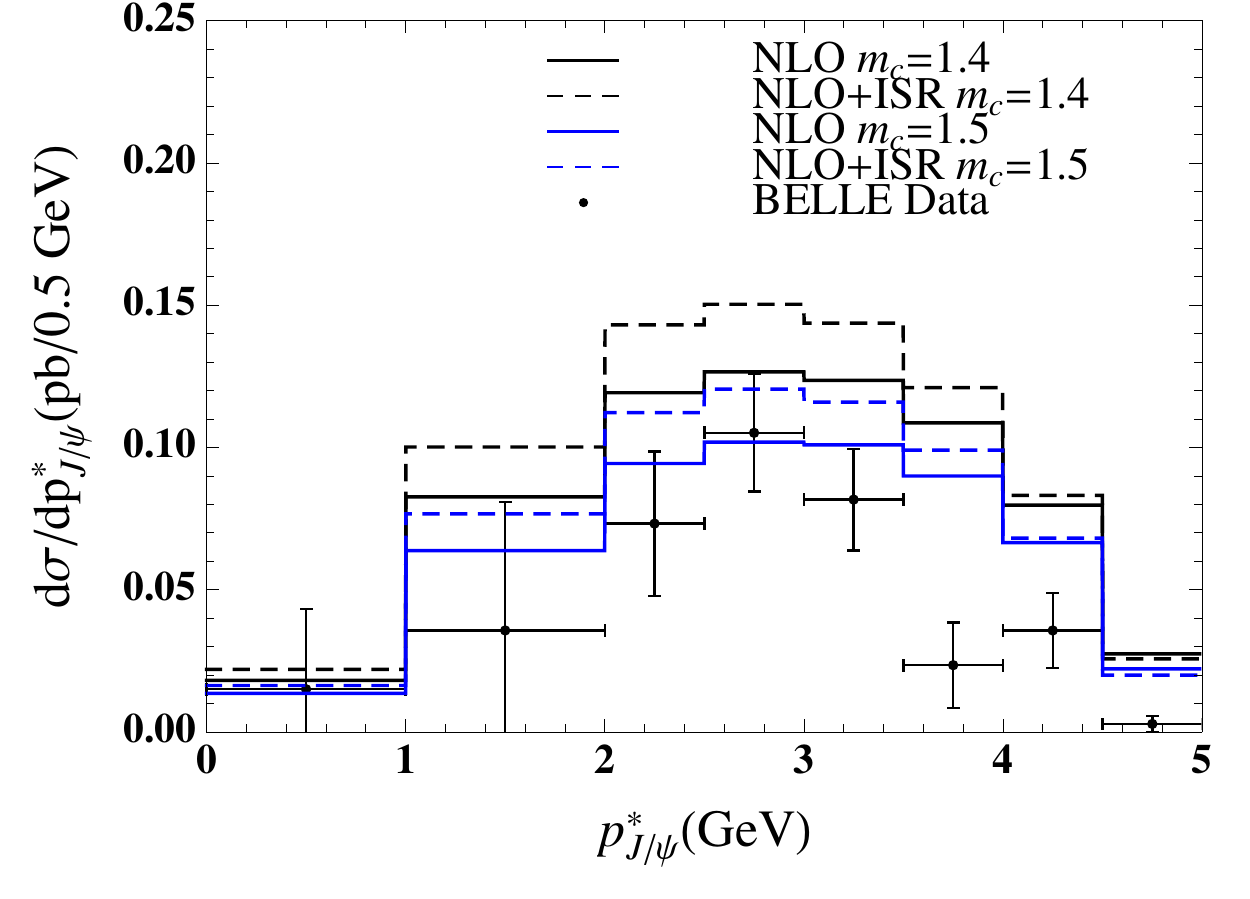}
\hspace{0cm}\includegraphics[width=0.45\textwidth]{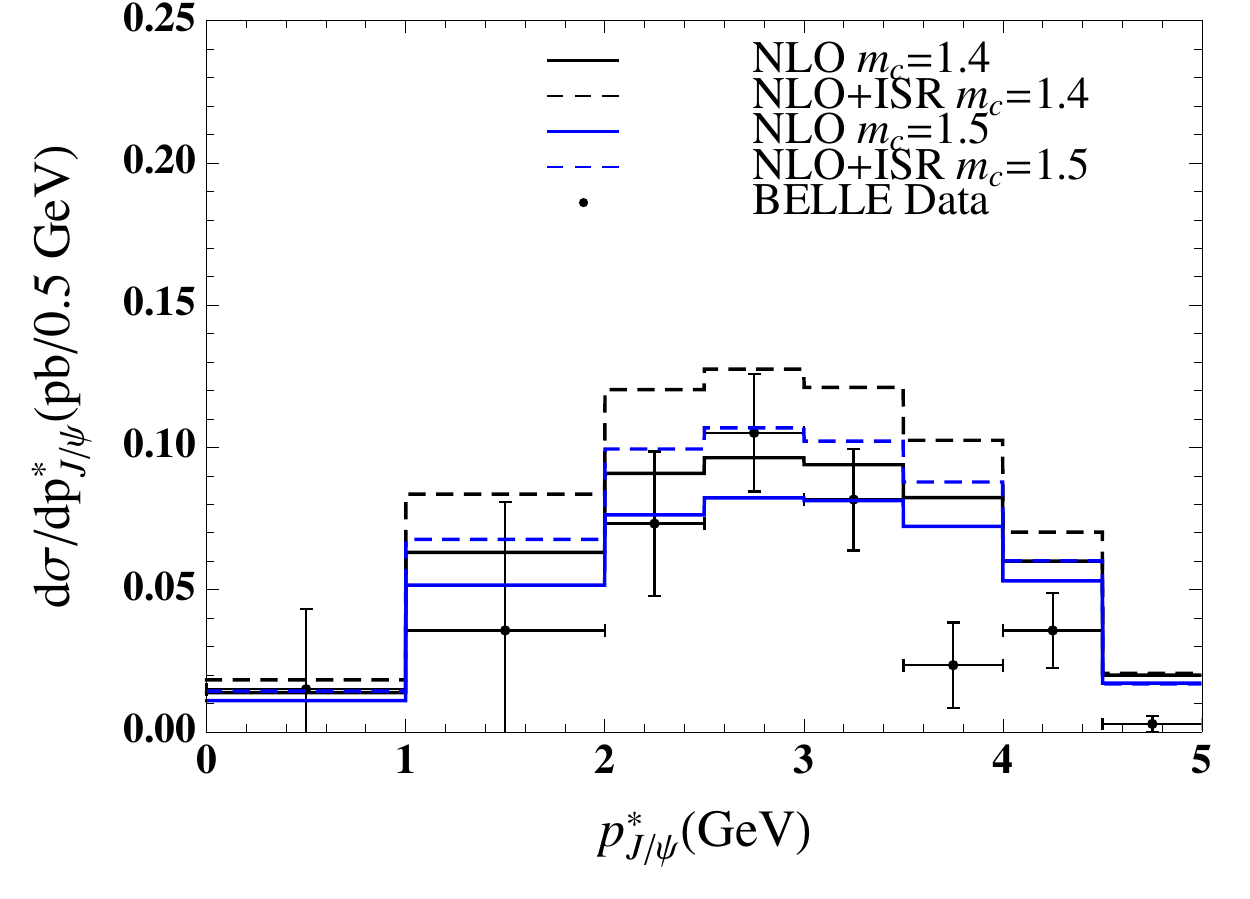}
\caption{\label{fig:dsigmadpBELLE}Comparisons of the theoretical predictions and \textbf{BELLE} measurement~\cite{Pakhlov:2009nj} with $\mu=2m_c$ (left-panel) and $\mu=\sqrt{\hat{s}}/2$ (right-panel) respectively in $e^-e^+\to \jpsi+gg+X$. We have multiplied a factor 1.355 to account for the feed-down contribution from $\psi(2S)\to \jpsi+X$.}
\end{figure}
\end{center}

\begin{table}
\begin{center}
\begin{tabular}{{c}*{4}{c}}\hline\hline
parameter sets & LO (pb) & LO+ISR (pb) & NLO (pb) & NLO+ISR (pb)\\\hline $m_c=1.4~\rm{GeV},\mu=2m_c$ & $0.57$ & $0.65$ & $0.79$ & $0.91$\\
$m_c=1.5~\rm{GeV},\mu=2m_c$ & $0.45$ & $0.50$ & $0.63$ & $0.72$\\
$m_c=1.4~\rm{GeV},\mu=\sqrt{\hat{s}}/2$ & $0.35$ & $0.45$ & $0.60$ & $0.77$ \\
$m_c=1.5~\rm{GeV},\mu=\sqrt{\hat{s}}/2$ & $0.30$ & $0.37$ & $0.51$ & $0.64$
\\\hline\hline
\end{tabular}
\end{center}
\caption{\label{tab:sigmagg}Cross sections of $e^-e^+\to \jpsi+gg+X$ in different parameter sets.}
\end{table}

\begin{table}
\begin{center}
\begin{tabular}{{c}*{5}{c}}\hline\hline
parameter sets & LO & LO+ISR & NLO & NLO+ISR & \textbf{BELLE}\\\hline $m_c=1.4~\rm{GeV},\mu=2m_c$ & $0.44$ & $0.41$ & $0.49$ & $0.45$ & $0.63\pm0.11$\\
$m_c=1.5~\rm{GeV},\mu=2m_c$ & $0.41$ & $0.38$ & $0.46$ & $0.42$ & $0.63\pm0.11$\\
$m_c=1.4~\rm{GeV},\mu=\sqrt{\hat{s}}/2$ & $0.47$ & $0.41$ & $0.50$ & $0.44$ & $0.63\pm0.11$\\
$m_c=1.5~\rm{GeV},\mu=\sqrt{\hat{s}}/2$ & $0.43$ & $0.38$ & $0.45$ & $0.40$ & $0.63\pm0.11$
\\\hline\hline
\end{tabular}
\end{center}
\caption{\label{tab:Rcc}Comparisons of $R_{c\bar{c}}$ in different parameter sets with \textbf{BELLE} measurement~\cite{Pakhlov:2009nj}.}
\end{table}

\begin{table}
\begin{center}
\begin{tabular}{{c}*{4}{c}}\hline\hline
parameter sets & LO (pb) & LO+ISR (pb) & NLO (pb) & NLO+ISR (pb)
\\\hline 
$m_c=1.4~\rm{GeV},\mu=2m_c$ & $1.02$ & $1.09$ & $1.55$ & $1.66$\\
$m_c=1.5~\rm{GeV},\mu=2m_c$ & $0.76$ & $0.80$ & $1.17$ & $1.25$\\
$m_c=1.4~\rm{GeV},\mu=\sqrt{\hat{s}}/2$ & $0.66$ & $0.76$ & $1.19$ & $1.36$\\
$m_c=1.5~\rm{GeV},\mu=\sqrt{\hat{s}}/2$ & $0.52$ & $0.59$ & $0.93$ & $1.05$
\\\hline\hline
\end{tabular}
\end{center}
\caption{\label{tab:crosssectiontot}Cross sections of $e^-e^+\to\jpsi+X$ in different parameter sets.}
\end{table}

For comparison with the \textbf{BELLE} measurement, we adapt the same bin size that they used. The $\jpsi$ momentum spectrum is shown in figure~\ref{fig:dsigmadpBELLE}. We obtain four different input parameter sets. The color-singlet result already saturaes the experimental data. With all of the three corrections (i.e. QCD,relativistic and ISR corrections), more stringent room is left for color-octet contribution in $\jpsi+X_{\rm{non-}c\bar{c}}$. The total theoretical cross sections for $e^-e^+\to\jpsi+gg+X$ in various parameter sets are summarized in table~\ref{tab:sigmagg}. ISR enlarges the cross section by $15-25\%$. Although the cross sections are slightly higher than the experimental data~\cite{Pakhlov:2009nj} and considering the large theoretical uncertainties, the theoretical result can still be lower. For example, we can use a lower value of $|R(0)|^2$ as performed in ref.\cite{Gong:2009kp} or from the potential model estimation~\cite{Eichten:1995ch}. In principle, the ratio $R_{c\bar{c}}$ should be independent of the value of $|R(0)|^2$ in color-singlet case. We present the theoretical $R_{c\bar{c}}$ in table~\ref{tab:Rcc}. We use a same parameter set in $\jpsi+c\bar{c}+X$ and $\jpsi+gg+X$ and assume that $\sigma(e^-e^+\to\jpsi+gg+X)=\sigma(e^-e^+\to\jpsi+X_{\rm{non-}c\bar{c}})$. The theoretical result is slightly lower than  \textbf{BELLE} measurement, but still within the $2$ standard deviation. Therefore, we expect that a more precise measurement could further elucidate the situation.Finally, we also list the total cross sections $\sigma(e^-e^+\to\jpsi+X)=\sigma(e^-e^+\to\jpsi+c\bar{c}+X)+\sigma(e^-e^+\to\jpsi+gg+X)$ in table~\ref{tab:crosssectiontot}, which is compatible with the experimental~\cite{Pakhlov:2009nj} value $\sigma_{\rm{prompt}}(e^-e^+\to\jpsi+X)=1.17\pm0.02\pm0.07~\rm{pb}$.

\section{Summary\label{sec:4}}
The different conclusions formulated from the large $p_T$ hadronic data and small scale $e^{\pm}$ data motivate us to reconsider the theoretical results. In this article, we focus on the issue about the prompt $\jpsi$ inclusive production at B factories. Considering that the cross sections in $e^-e^+$ annihilation usually gain large corrections from ISR, we use the MC techniques to include the ISR effect in our theoretical results. Results indicate that the effect of ISR photon shower to $e^-e^+\to\jpsi+c\bar{c}+X$ is small but it is large to $e^-e^+\to\jpsi+gg+X$. ISR enhances the total cross section of  $e^-e^+\to\jpsi+gg+X$ by a factor of $15-25\%$, which depends on the input values of the parameters. Moreover, ISR causes the $\jpsi$ momentum spectrum to be softer in $e^-e^+\to\jpsi+c\bar{c}+X$ and $e^-e^+\to\jpsi+gg+X$. This result is important because it provides a good perspective to the color-octet contribution. We present the theoretical results for these two processes in combination with the QCD,relativistic, and ISR corrections. The feed-down contributions (mainly from $\psi(2S)$) are also included for comparison with the experimental data. The total cross sections for $\jpsi+c\bar{c}+X$ and $\jpsi+gg+X$ are presented in table~\ref{tab:sigmacc} and table~\ref{tab:sigmagg} respectively, whereas the $\jpsi$ momentum spectra in the \textbf{BELLE} bin size~\cite{Pakhlov:2009nj}  are shown in figure~\ref{fig:dsigmadpBELLEcc} and figure~\ref{fig:dsigmadpBELLE}. Considering the large experimental and theoretical uncertainties, we are still unable to draw strong conclusions. However, the corrections in $\jpsi+gg+X$ considerably constrain the color-octet contribution. Finally, we also present $R_{c\bar{c}}$ with the assumption $\sigma(e^-e^+\to\jpsi+gg+X)=\sigma(e^-e^+\to\jpsi+X_{\rm{non-}c\bar{c}})$, which should be more precise than the cross section alone. The theoretical result is lower than \textbf{BELLE} measurement~\cite{Pakhlov:2009nj}, but still within $2\sigma$. More careful theoretical and experimental analyses are necessary in the future.

\begin{acknowledgments}
We are thankful to Prof.Kuang-Ta Chao, Dr. Yu-Jie Zhang and Dr. Zhi-Guo He for helpful discussions. 
This work was supported in part by the National Natural Science
Foundation of China (No 11075002, No 11021092), and the Ministry of
Science and Technology of China (2009CB825200).
\end{acknowledgments}




\providecommand{\href}[2]{#2}\begingroup\raggedright\endgroup

\end{document}